\documentclass[preprint2]{aastex}
\usepackage{epsfig}

\newcommand{\der}[2]  { \frac{{\rm d}#1}{{\rm d}#2} }

\newcommand{\dif}     {{\rm d}}

\shorttitle{Supegalactic winds} 
\shortauthors{} 
 
\begin{document} 

\title
{On the rapidly cooling interior of supergalactic winds}

\author{S. Silich \altaffilmark{1,2}, 
        G. Tenorio-Tagle \altaffilmark{1} and 
        C. Mu\~noz-Tu\~n\'on \altaffilmark{3}}
\altaffiltext{1}{Instituto Nacional de Astrof\'\i sica Optica y Electr\'onica, 
                 AP 51, 72000 Puebla, M\'exico}
\altaffiltext{2}{Main Astronomical Observatory National Academy of Sciences of
                 Ukraine, 03680, Kiev-127, Golosiiv, Ukraine}
\altaffiltext{3}{Instituto de Astrofisica de Canarias, E 38200 La Laguna, 
                 Tenerife, Spain}

\begin{abstract} 
Here we present the steady state numerical solution and two
dimensional hydrodynamic calculations of supergalactic winds, 
taking into consideration strong radiative cooling. The two possible 
outcomes: quasi-adiabatic and strongly radiative flows
are thoroughly discussed, together with their implications 
on the appearance of supergalactic winds both in the X-ray and visible
line regimes.
\end{abstract} 
 

\section {Introduction} 
 
Massive starbursts, well localized short episodes of violent star 
formation, are among the intrinsic characteristics of the evolution 
of galaxies. They are found both at high and intermediate  redshifts 
(Dawson et al., 2002; Ajiki et al., 2002; Pettini et al., 1998) as well as  
in galaxies in the local universe (Marlowe et al., 1995, 
Cairos et al., 2001).

It has usually been assumed that the energy deposition occurs within
a region of about 100 pc, the typical size of a starburst. However
recent optical, radio continuum and IR observations (Ho 1997; 
Johnson et al. 2001, Gorjian et al. 2001) revealed a number of 
unusually compact young stellar clusters.
These overwhelmingly luminous concentrations of stars present a
typical radius of about 3 pc, and a mass that ranges from a few 
times 10$^4$M$_{\odot}$ to 10$^6$M$_\odot$. Clearly these units
of star formation (super-star clusters) are very different from what
was usually assumed to be a typical starburst, and we shall put a 
special emphasis on the outflows expected from them.

The large kinetic luminosity produced by violent bursts of star 
formation is well known to drastically affect the surrounding
interstellar medium, generating the giant superbubbles and
supershells, detected  in a large sample of star-forming galaxies
(Bomans, 2001).  
In other cases, the energy input rate is able to drive their 
associated shock waves to reach the outskirts of their
host galaxy, leading, in these cases, to the development of a
supergalactic wind with outflow rates of up to several tens of solar
masses per year (see Heckman 2001 for a complete review). One of the 
main implications of such outflows 
is the clear contamination of the intergalactic medium with the metals 
recently processed by the massive starburst. In both cases it is
usually assumed that at the heart of the starburst, within the region 
that encompasses the recently formed stellar cluster (R$_{SB}$),
the matter ejected by strong stellar winds and supernova explosions is 
fully thermalized. This generates the large central overpressure
responsible for the mechanical luminosity associated with the new starburst.  
Within the starburst region, it is  the mean total mechanical 
energy L$_{SB}$ and mass ${\dot M}_{SB}$ deposition rates that
control, together with the actual size of the star forming region 
R$_{SB}$, the properties of the resultant outflow. The total mass 
and energy deposition rates define the central temperature T$_{SB}$
and thus the sound speed c$_{SB}$ at the cluster boundary. 
As shown by Chevalier \& Clegg (1985; hereafter referred to as CC85) 
at the boundary r = R$_{SB}$, the thermal and kinetic energy flux
amount exactly to 9/20 and 1/4 of the total energy flux: 
\begin{eqnarray}
      \label{eq.02} 
      & & \hspace{-0.5cm}
F_{th}/F_{tot} = \frac{\frac{1}{\gamma - 1}\frac{P}{\rho}}
                 {\frac{u^2}{2} + \frac{\gamma}{\gamma - 1}
                 \frac{P}{\rho}} = \frac{9}{20} , 
      \\[0.2cm]
      & & \hspace{-0.5cm}
F_{k}/F_{tot} = \frac{u^2/2}{\frac{u^2}{2} + \frac{\gamma}{\gamma - 1}
                 \frac{P}{\rho}} = \frac{1}{4} .
\end{eqnarray}
There is however a rapid evolution as matter streams away from the
central starburst. After crossing $r = R_{SB}$ the gas is immediately 
accelerated by the steep pressure gradients and rapidly reaches 
its terminal velocity (V$_{\infty} \approx 2$ c$_{SB}$). This is due to a fast 
conversion of thermal energy, into kinetic energy in the resultant 
wind. In this way, as the wind takes distance to the star cluster
boundary, its density, temperature and thermal pressure will
drop as r$^{-2}$, r$^{-4/3}$ and  r$^{-10/3}$, respectively (CC85). 
Previous analysis by Heckman et al. (1990) showed that thermal pressure 
radial distributions inside starburst galaxies M82 and NGC 3256 
are in a good agreement with the CC85 model. However Martin et
al. (2002) found that X-ray surface brightness and radial
temperature gradient observed by Chandra in NGC 1569 are inconsistent
with CC85 predictions.

Note that the wind is exposed to the appearance of reverse 
shocks whenever it meets an obstacle cloud or when its thermal pressure   
becomes lower than that of the surrounding gas, as is the case within
superbubbles. There, the high pressure acquired by the swept up ISM 
becomes larger than that of the freely expanding "free wind region" (FWR),
a situation that rapidly leads to the development of a reverse shock,
the thermalization of the wind kinetic energy  
and to a much reduced size of the FWR. Thus for the FWR to extend 
up to large distances away from the host galaxy, the shocks  would have had
to evolve and displace all the ISM, leading to a free path into the 
intergalactic medium and to a supergalactic wind
with properties similar to those derived by CC85. 
Here we study the true physical properties of such well developed
FWRs, taking into consideration strong radiative cooling. Section 2  
compares the adiabatic solution of CC85 with our steady state solution 
where radiative cooling is considered. In section 2 we also develop
an easy criterion to define in which cases radiative cooling is to 
become dominant. Section 3 displays two dimensional calculations 
that use CC85 as the initial condition which evolves until a new steady state 
is reached. Section 4 discusses some of the observational consequences 
of such outflows.

\section {The steady state solution}

Following CC85 we assume a spherically symmetric 
wind, unaffected by the gravitational pull caused by the central star 
cluster or its associated dark matter component. The equations that 
govern the steady outflow away from  the star forming region are:
\begin{eqnarray}
      \label{eq.1a}
      & & \hspace{-0.5cm}
\frac{1}{r^2} \der{}{r}\left(\rho u r^2\right) = 0 ,
      \\[0.2cm]
      \label{eq.1b}
      & & \hspace{-0.5cm}
\rho u \der{u}{r} = - \der{P}{r} ,
      \\[0.2cm]
     \label{eq.1c}
      & & \hspace{-0.5cm}
\frac{1}{r^2} \der{}{r}{\left[\rho u r^2 \left(\frac{u^2}{2} +
\frac{\gamma}{\gamma - 1} \frac{P}{\rho}\right)\right]} = - Q,
\end{eqnarray}
where $r$ is the spherical radius, $u(r), \rho(r)$ and $P(r)$ are 
the wind velocity, density and thermal pressure, respectively. $Q$ is
the cooling rate  ($Q = n^2 \Lambda$, where $n$ is the wind number 
density and $\Lambda$ is the interstellar cooling function, Raymond et
al. 1976). At the boundary of the star cluster ($R_{SB}$) we use 
the solution of CC85
\begin{eqnarray}
      \label{eq.2a}
      & & \hspace{-0.5cm}
{\dot M_{SB}} = (0.1352 \pi \gamma)^2 \frac{L_{SB}}{c_{SB}^2} ,
      \\[0.2cm] \label{eq.2b}
      & & \hspace{-0.5cm}
P = 0.0338 {\dot M_{SB}}^{1/2} L_{SB}^{1/2} R_{SB}^{-2} ,
      \\[0.2cm] \label{eq.2c}
      & & \hspace{-0.5cm}
T = \frac{0.299 \mu}{k} \frac{L_{SB}}{{\dot M}_{SB}} , 
      \\[0.2cm] \label{eq.2d}
      & & \hspace{-0.5cm}
u = c_{SB} = \left(\frac{\gamma k T}{\mu}\right)^{1/2} , 
      \\[0.2cm] \label{eq.2e}
      & & \hspace{-0.5cm}
\rho = \frac{{\dot M_{SB}}}{4 \pi R_{SB}^2 c_{SB}} ,
\end{eqnarray}
where $\mu$ is the mean mass per wind particle, $k$ is the Boltzmann 
constant, $\gamma = 5/3$ is the ratio of specific heats and c$_{SB}$ 
is the sound speed at r = R$_{SB}$, and then solve equations 
(\ref{eq.1a})-(\ref{eq.1c}) numerically. Note that at r = R$_{SB}$ 
the wind Mach number M$_w = 1$. Therefore we use the wind velocity 
as an independent variable at the vicinity of the sonic point. 
Our numerical solutions for $Q = 0$, fully reproduce CC85 adiabatic 
results.

\subsection {Radiative cooling in supergalactic winds}

A first order of magnitude estimate of whether or not radiative cooling 
could affect the thermodynamics of superwinds, results  from a 
comparison of the radiative cooling time scale 

\begin{equation}
      \label{eq.4} 
\tau_{cool}(r) = \frac{3 k T}{n \Lambda} ,
\end{equation} 
with the characteristic dynamical time scale  
\begin{equation}
      \label{eq.5} 
\tau_{dyn}(r) = \int_{R_{SB}}^r \frac{\dif{r}}{u(r)} .
\end{equation} 
The wind density, which, as it streams away, decreases as (1/$r^2$), in 
combination with the interstellar cooling curve, fully define locally 
the ratio of $\tau_{cool}/\tau_{dyn}$. One can infer then that if the 
ratio of the above two quantities becomes smaller than 1, radiative 
cooling sets in, affecting the thermodynamical properties of the flow. 
These simple estimates indicate that radiative cooling may become more 
efficient the more compact a star cluster is. In fact, it is the 
density of the free wind what really matters. For the same total
starburst energy, the more compact a starburst is, the larger 
the wind density value and thus the larger the cooling rate.  
The temperature profiles for a $10^{41}$ erg s$^{-1}$ starburst
with a radius R$_{SB} = 5$ pc and different metallicities of the 
wind material are shown in Figure 1, for both the adiabatic and the 
radiative cases. Note that for compact starbursts radiative cooling 
sets in within a radius $\leq$ 10 $R_{SB}$. Despite the rapid drop 
in temperature the velocity and the density distributions
remain almost unaffected by radiative cooling. In all cases  
the thermal energy is rapidly transformed into kinetic energy of the
wind in the close vicinity of the sonic point. This leads to a fast
gas acceleration that allows the wind gas to rapidly approach its
terminal velocity V$_{\infty} \approx 2$ c$_{SB}$. At the same time, if 
radiative cooling is considered, the temperature profile strongly 
deviates away from the adiabatic solution forcing the gas to soon 
reach temperature values of the order of 10$^4$ K.
 
One can establish an approximate boundary that separates the adiabatic
from the strongly radiative cases, from the condition that 
$\tau_{cool}$ becomes equal to $\tau_{dyn}$. The corresponding curves 
for different energy input rates, wind terminal velocities and wind 
metallicities, as a 
function of R$_{SB}$, are shown in Figure 2. If the initial wind 
parameters (L$_{SB}$ and $R_{SB}$) intersect below the corresponding 
metallicity curve, cooling will be inefficient  and deviations 
from the adiabatic solution would be negligible. On the other hand, 
if the initial wind parameters intersect above the corresponding  
metallicity curve, radiative cooling is expected to become important.
Note that the larger the wind terminal velocity is, the smaller the region
of the parameter space covered by the cold wind solution.

\section{The time dependent solution}

Several two dimensional calculations using CC85 as the initial condition 
adiabatic flows have been performed with the Eulerian code described 
by Tenorio-Tagle \& Mu\~noz Tu\~non (1997, 1998). This has been
adapted to allow for the continuous injection in the most central
zones, of winds with a variety of L$_{SB}$ and ${\dot M}_{SB}$
values. The time dependent calculations also account for radiative 
cooling and display the evolution from the adiabatic regime to the 
more realistic final steady state solution when radiative cooling is 
considered. There is an excellent agreement between the time dependent 
calculations and the steady state solutions displayed in section 2.

Figure 3 displays isotemperature contours and the velocity field within 
the central 100 pc$^2$ of a supergalactic wind driven by an 
L$_{SB} = 10^{42}$ erg s$^{-1}$, c$_{SB}$ = 500 km s$^{-1}$ or a 
V$_{\infty}$ = 1000 km s$^{-1}$, emanating from an  R$_{SB} = 20$ pc.
The solution for Z = Z$_{\odot}$, as expected from Figure 2, is well 
into the strongly radiative regime. Figure 3 shows the rapid increase 
in velocity from its central value to the terminal speed $\sim 2$ c$_{SB}$. 
The upper panel shows the run of temperature for the corresponding
CC85 adiabatic solution, our initial condition, while the central
panels depict how radiative cooling proceeds within the supersonic 
outflow, leading to the formation of cool parcels of gas 
(T $\sim$ 10$^4$ K) within rapidly expanding shells of cooling gas.
After a few 10$^4$ years a new steady state solution has been reached 
(lower panel). There the temperature of the fast moving wind suddenly 
plummets from 10$^7$ K to 10$^4$ K at a distance of 37 pc from $R_{SB}$.
Cooling takes place in a catastrophic manner, a fact that made us
search for possible signs of fragmentation, however, the sudden 
pressure gradients resulting from cooling are inhibited by the 
strongly diverging flow, leaving both the density and the velocity 
fields almost unperturbed. Similar results were obtained when  higher 
metallicity outflows were assumed. The high metallicities are indeed 
expected from the large inflow of new metals into the superwind 
(see Silich et al., 2001). Figure 4 shows the final steady-state 
solution for a wind with R$_{SB} = 5$ pc, L$_{SB} = 10^{41}$ erg
s$^{-1}$ and V$_{\infty} = 1000$ km s$^{-1}$ when the assumed metallicity 
of the wind is equal to 1$Z_\odot$, 3$Z_\odot$, 5$Z_\odot$ and 
10$Z_\odot$. Clearly the impact of this variable is to favor 
radiative cooling even closer to the R$_{SB}$ surface, fact that
can be also noticed in Figure 1.

\section{The structure of supergalactic winds}

The results from sections 2 and 3 
imply that the structure of supergalactic winds aught to be
revised, particularly in the case of  powerful sources with a 
small value of R$_{SB}$ in which radiative cooling reduces the spatial 
extent of the X-ray emitting region. In such cases, the only
possibility for the origin of an extended hot gas component arises 
from shock heated ISM overrun by the central wind. On the other 
hand, in lower luminosity and/or widely spread starbursts, the extended
hot wind region may also contribute significantly to the total soft 
X-ray emission. Here we center our attention on very massive and 
centrally concentrated starbursts, entities that leave no doubt that 
their newly processed elements will sooner or later be driven into 
the IGM. In such cases the central wind is affected by radiative 
cooling and the extended structure of such winds will be undetected 
in the X-ray regime. 
 
Supergalactic winds present a four zone structure: 1) A central
starburst region (a source of hard and soft X-rays). 2) A soft X-ray 
emitting zone. 3) The line cooling zone. 4) A region of recombined 
gas, exposed to the UV radiation from the central star cluster.

Figure 5 displays the radial structure of supergalactic winds powered 
by low (L$_{SB} = 10^{41}$ erg s$^{-1}$) and high 
(L$_{SB} = 10^{43}$ erg s$^{-1}$) luminosity starbursts as a function
of R$_{SB}$. In all cases we have assumed a free wind region terminal 
velocity equal to 1000 km s$^{-1}$ and a 1Z$_\odot$ metal abundance. 
Solid lines mark the distance at which a $10^{41}$erg s$^{-1}$
superwind acquires a temperature of $5 \times 10^5$K and $10^4$K  
for an adiabatic solution. Dashed lines present the modified location 
of the two temperature boundaries provided by gas cooling. Dotted 
lines mark the position of the two temperature boundaries for energetic
starbursts ($10^{43}$erg s$^{-1}$).
Note that in both high and low luminosity cases we have 
assumed the same central temperature, and thus the same central sound 
velocity (c$_{SB}$), both cases develop the same terminal speed,
leading in the adiabatic regime to the same temperature radial 
structure. For the low luminosity 
radiative cases (heavy dashed lines) significant departures from the 
adiabatic solution occur for highly concentrated starbursts 
(say $R_{SB} \leq$ 20 pc) bringing the zone of rapid radiative cooling 
($10^4$ K $\le$ T $\le$ 5 $\times 10^5$ K) closer to the 
starburst surface. On the other hand, the radiative solution for the 
energetic cases (dotted lines) shows how both temperature limits here 
considered move much closer to the starbursts, reducing significantly the 
extent of the X-ray and the line cooling zones. The effect is noticeable 
for all $R_{SB}$ (from 1pc to 100 pc) values here considered, a fact
that ought to be taken into account in full numerical simulations.

\subsection{The impact of photo-ionization and the expected Halpha luminosity} 

Figure 5 shows that powerful and compact starbursts are to be 
surrounded by extended, cold and fast expanding envelopes. This 
continuously injected halo, however, becomes an easy target of the 
UV photons escaping from the central star forming regions, and as it 
presents an  $R^{-2}$  density distribution it should be  completely 
ionized by the stellar radiation. This is simply because
the number of recombinations in such a density distribution is, as
shown by  Franco et al. 1990, unable to trap the ionization front.
As shown in Figure 5, the size of the X-ray and line cooling zones
become strongly affected by radiative cooling reducing largely 
their dimensions in favor of a closer development of the outer 
photo-ionized region 4. Note that the recombining zone 3 is also 
an additional source of UV photons capable of ionizing the wind 
outer extended halo. The number of recombinations per unit time 
expected in the photo-ionized, outer free wind region is:

\newpage

\begin{eqnarray}
      \label{eq.8}
      & & \hspace{-1.2cm} \nonumber
\der{N_{rec}}{t} = 4 \pi R_4^2 n(R_4) V_w(R_4) + 
      \\[0.2cm] \label{eq.2d}
      & & \hspace{-1.2cm}
4 \pi \beta \int_{R_4}^{\infty} n^2(r) r^2 \dif r =
    \frac{{\dot M}_w}{\mu_a} + \frac{\beta {\dot M}_w^2}
    {4 \pi V_t^2 \mu_a^2 R_4} ,
\end{eqnarray}
where the first term in the right hand side refers to the number of 
recombinations in the rapidly cooling zone and the second one is
related to the number of recombinations in the outer cold envelope (zone 4).
Recombinations should lead to H$\alpha$ and Ly$\alpha$ emission. One
then can estimate the H$_{\alpha}$ luminosity L$_{H\alpha} = 
1.36 \times 10^{-12} {\dot N_{rec}}$ (Leitherer \& Heckman, 1995) 
and Ly$\alpha$ luminosity L$_{Ly\alpha} \approx 8.74$ $L_{H\alpha}$
(Brocklehurst, 1971). The expected H$_{\alpha}$ luminosities
calculated for radiative winds powered by $10^{41}$ erg s$^{-1}$ and  
$10^{43}$ erg s$^{-1}$ starbursts are presented in Figure 6 as
a function of R$_{SB}$.

\subsection{The X-ray luminosity}

The X-ray luminosity from the hot wind with density distribution
$n(r) = {\dot M}_w / 4 \pi \mu_a u(r) r^2$ is 
\begin{equation}
      \label{eq.10} 
L_x = 4 \pi \int_{R_{SB}}^{R_{x}} n^2 \Lambda_x(T,Z_w) r^2 
      \dif r ,
\end{equation} 
where $\Lambda_x(T,Z_w)$ is the hot gas X-ray emissivity and R$_x$ 
is the X-ray emission cutoff radius that corresponds to the cutoff
temperature T$_{cut} \approx 5 \times 10^5$K. This is shown 
in Figure 5. We then use a model velocity field u(r) and the 
cutoff radius R$_x$ to calculate a wind soft (0.1kev - 2.4kev) 
X-ray luminosity. The results of the calculations
for L$_w = 10^{41}$ erg s$^{-1}$ and L$_w = 10^{43}$ erg s$^{-1}$
superwinds are shown in Figure 6. We have to conclude that the 
inconsistency between the free wind model predictions and observed 
superwind X-ray luminosities, mentioned by Strickland \&  Stevens
(2000), becomes even larger if one takes into account the modifications 
provided by radiative cooling. That is, in the standard approach
the free wind material itself cannot be the main source of diffuse
superwind X-ray emission.

\section{Conclusions}

We have reanalyzed the physical properties (velocity, density and 
temperature) of supergalactic winds powered by a constant energy 
input rate, taking radiative cooling into account. Three input 
parameters control the properties of the wind: the total energy 
input rate L$_{SB}$, the characteristic scale of star formation
(R$_{SB}$), and a thermalized gas temperature T$_{SB}$ (or initial wind 
velocity c$_{SB}$). We have revealed two possible outflow regimes:
a hot wind, which due to its low power and low density remains
unaffected by cooling, and thus as found by CC85 behaves
adiabatically. On the other hand, winds driven by  powerful and/or 
compact starbursts become rapidly dominated by  
catastrophic radiative cooling. 

Superwinds driven by compact and powerful starbursts (see Figure 2)
undergo catastrophic cooling close to their sources,  
and establish a temperature distribution radically different 
from those obtained in  adiabatic calculations. At the same time, both
velocity and density radial distributions remain almost unchanged 
given the speed of the rapidly diverging flow. 

The rapid fall in temperature as a function of r 
reduces the size of the zone radiating in X-rays and decreases 
the superwind X-ray luminosity. At the same time, cooling brings the 
10$^4$ K boundary closer to the wind center, and promotes the 
establishment of an extended ionized fast moving envelope.
This should show up as a weak and broad ($\sim 1000$ km s$^{-1}$) 
line emission component at the base of the much narrower line caused 
by the central HII region.

The authors highly appreciated the friendly hospitality of A. D'Ercole,
F. Brighenti, M. Tosi and F. Matteucci at the superwind  2002 workshop
in Bologna, where this study was initiated. They thank the referee for
helpful comments and Jeff Wagg for his careful reading of the manuscript.
This study has been  supported by  CONACYT - M\'exico, 
research grant 36132-E and by the Spanish Consejo Superior de 
Investigaciones Cient\'\i{}ficas, grant AYA2001 - 3939.

\clearpage 
 
\figcaption[fig1.ps] 
{The intrinsic temperature profile of superwinds. The dash-dotted 
line shows the temperature distribution derived from the 
adiabatic solution of CC85. Other curves show the effects 
of radiative cooling for different superwind metallicities.
\label{fig1}} 
 
\figcaption[fig2.ps] 
{Adiabatic versus radiative superwinds. The lines divide the
region defined by the parameter space into strongly radiative 
superwinds (above the lines) and adiabatic solutions (below
the lines), for different values of the terminal velocity 
(V$_{\infty}$) and metallicity.
\label{fig2}}

\figcaption[fig3.ps] 
{Two dimensional superwinds. Isotemperature contours, with a
separation $\Delta \log$T = 0.1 and the velocity field 
(longest arrow = 1000 km s$^{-1}$) within the central 100 pc$^2$
of a superwind powered by a $10^{42}$ erg s$^{-1}$ starburst with
a radius R$_{SB} = 20$pc. The evolution starts from (a) the adiabatic
solution of CC85 and continues at (b) t = $1.3 \times 10^4$yr,
(c) $1.6 \times 10^4$yr and (d) $1.9 \times 10^4$yr 
when a new steady state solution is reached. In the bottom 
panel the gas temperature falls to $10^4$K before reaching 40pc 
from the center. Distance between consecutive tick mark = 25pc in 
all figures.
\label{fig3}}

\figcaption[fig4.ps] 
{The metallicity effects. The same as figure 3 for a $10^{41}$erg 
s$^{-1}$ superwind with an R$_{SB} = 5$pc. The panels show the 
final temperature and velocity distributions for the radiative steady
state solution for (a) Z = Z$_{\odot}$, (b) 3Z$_{\odot}$,
(c) 5Z$_{\odot}$ and (d) 10Z$_{\odot}$. 
\label{fig4}}

\figcaption[fig5.ps] 
{The internal structure of free wind regions. Solid lines mark
the distance at which a $10^{41}$erg s$^{-1}$ superwind, with
a velocity V$_{\infty} = 1000$km s$^{-1}$, acquires a temperature
of $5 \times 10^5$K and $10^4$K as a function of the initial radius
R$_{SB}$, for an adiabatic solution. Dashed lines present the
modified location of the two temperature boundaries
provided by gas cooling (Z = Z$_{\odot}$). Dotted lines mark
the position of the two temperature boundaries for energetic
starbursts ($10^{43}$erg s$^{-1}$).
\label{fig5}}

\figcaption[fig6.ps] 
{Observational appearance of free wind regions. The expected X-ray
and H$_{\alpha}$ luminosity of superwinds as a function of the
starburst power and the radius R$_{SB}$.
\label{fig6}}

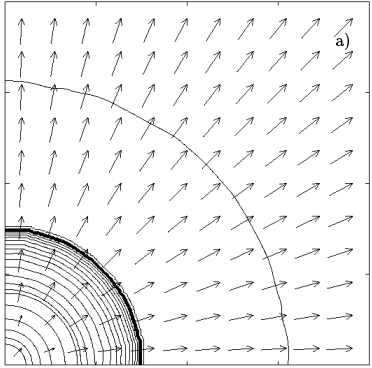
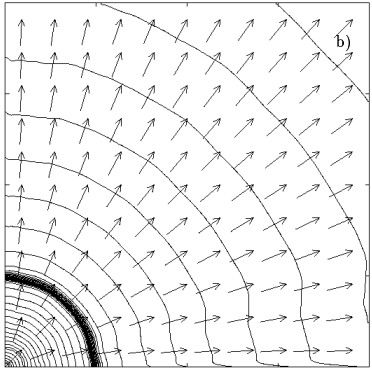
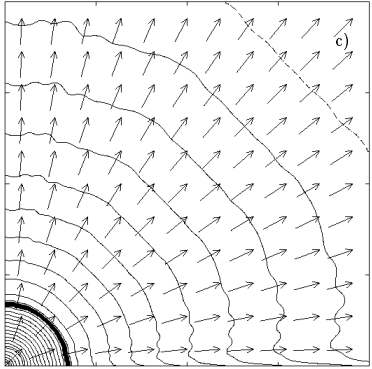
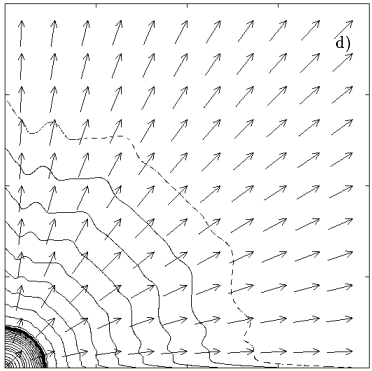
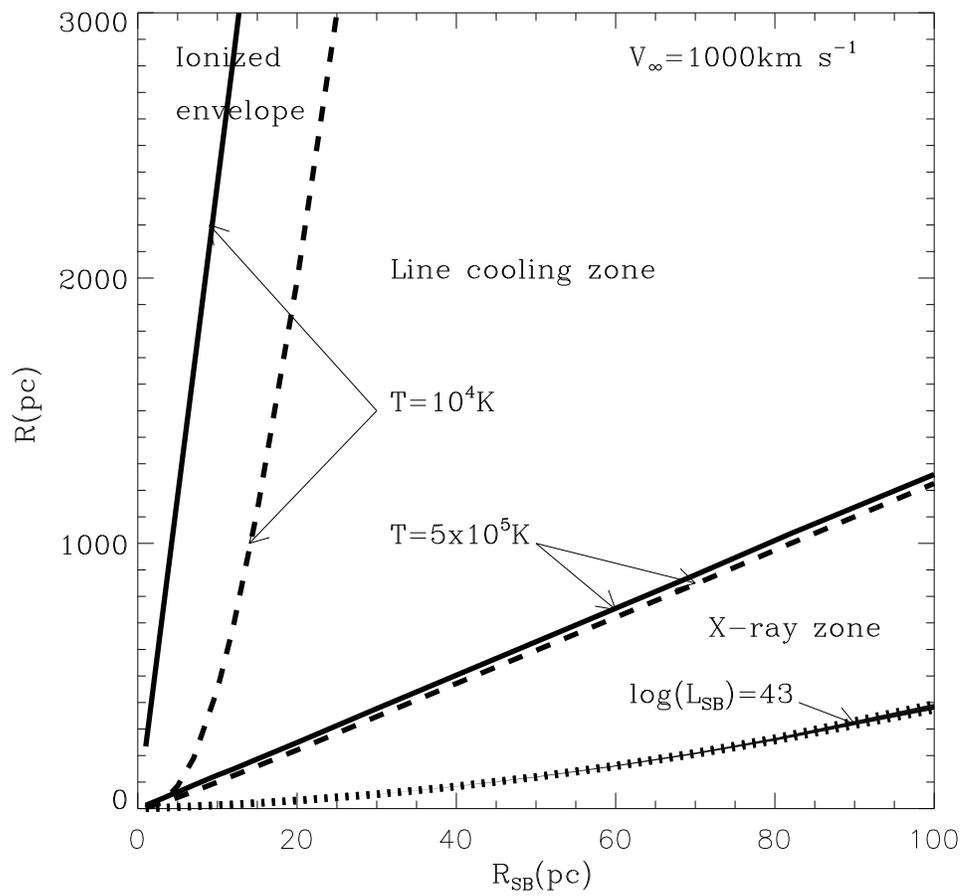
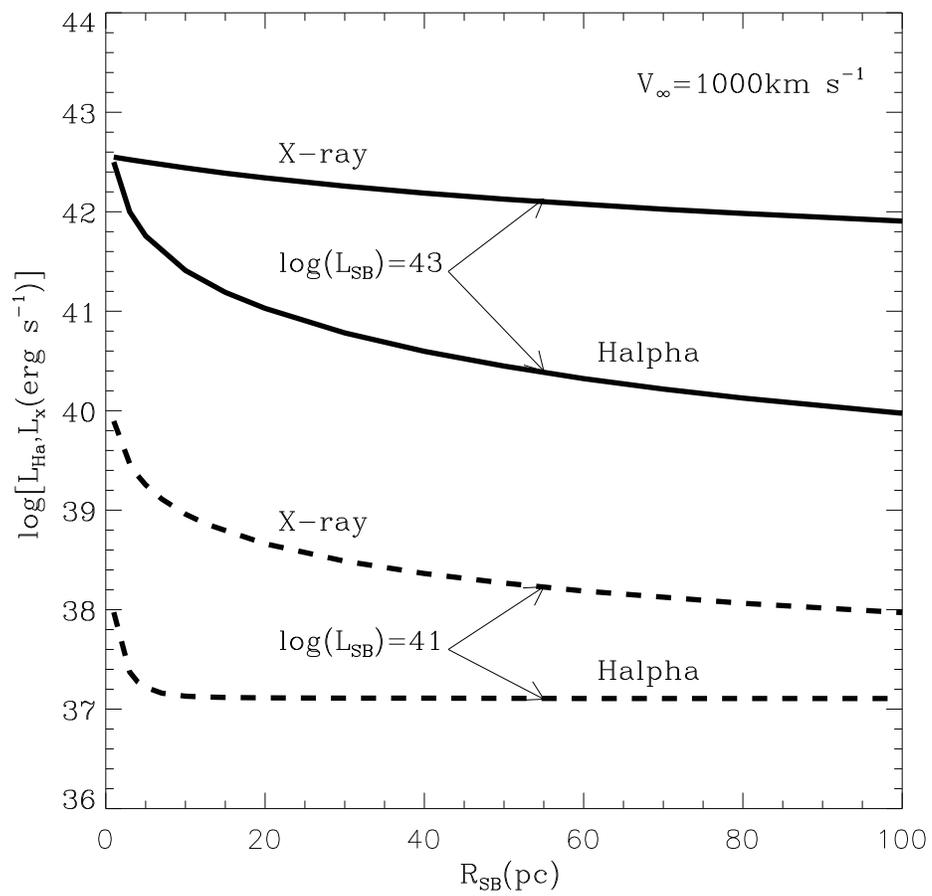
\end{document}